\documentclass[doublecol]{epl2}
\usepackage{amsmath}
\title{Log-periodic modulation in one-dimensional\\ random walks}
\shorttitle{Log-periodic modulation in 1D RW} 

\author{L.~Padilla \and H.~O.~M\'artin \and J.~L.~Iguain}
\shortauthor{L.~Padilla\etal}

\institute{                   
Instituto de Investigaciones F\'{\i}sicas de Mar del Plata (IFIMAR) and 
Departamento de F\'{\i}sica\\
 FCEyN, Universidad Nacional de Mar del Plata,
De\'an Funes 3350, (7600) Mar del Plata, Argentina.\\
}

\abstract{We have studied the diffusion of a single particle on a 
one-dimensional lattice. It is shown that, for a self-similar distribution of 
hopping rates, the time dependence of the mean-square displacement follows
an anomalous power law modulated by logarithmic periodic oscillations.
The origin of this modulation is traced to the dependence on the length of the
diffusion coefficient.
Both the random walk exponent and the
period of the modulation are analytically calculated and confirmed by
Monte Carlo simulations.
}

\begin{document}

\maketitle

Brownian Motion is a well-known phenomenon, and since its theoretical 
foundations were laid, more than one hundred years ago\cite{ein,smo},
diffusion processes and random walk models have been attracting the attention
of researchers.  The importance of random walk (RW) resides in the fact that it
is the simplest realisation of Brownian Motion, with applications in almost
 every field of science where stochastic dynamics play a role\cite{hug}.
It is worth to remark that, even though a random walker evolves according
to simple rules, a considerable effort may be needed to solve
the dynamic problem in detail and unexpected behaviours may emerge.
Thus, for example, our understanding
of the mechanisms responsible for anomalous diffusion has been strongly
fluenced by the large amount of work  devoted to the 
study of RW in non-Euclidean media, during the last
three decades\cite{ale,ram,ben}.

In the last years, it has been often reported that, sometimes, 
the time behaviour of 
a RW is modulated by logarithmic-periodic oscillations. 
These fluctuations has been
rigorously studied by mathematicians  on special kinds of 
graphs. A proof of the fluctuating behaviour
 of the $n$-step probabilities for a simple RW on a Sierpi\'nski graph was 
given in ref.~\cite{gra} and a 
generalisation  to the broad class of symmetrically self-similar 
graphs can be found in ref.~\cite{kro}.
Within the physical community, it has been shown that, on Sierpi\'nski gaskets,
 the mean number of distinct sites visited at time {\it t} by {\it N}
noninteracting random walkers presents an
oscillatory behaviour \cite{ace} 
and, more recently, detailed studies of the log-periodic
modulations on fractals with finite ramification order, were presented in
refs.~\cite{bab1,bab2}.

Log-periodic modulations are not restricted to random walks. It is in 
general believed that they appear because of an inherent 
self-similarity\cite{dou},
responsible for a
discrete scale invariance (DSI)\cite{sor1}.
Examples of these oscillations have been detected in 
earthquakes \cite{huang,saleur}, 
escape probabilities in chaotic maps close to crisis\cite{pola}, biased
diffusion\cite{stau,yu}, kinetic and 
dynamic processes on random quenched and fractal 
media\cite{bernas,kut,andra,bab3}, and 
stock markets near a financial crash\cite{sor2,van1,van2,van3}.

In this work
we analyse a minimal model of RW, which results in log-periodic modulations
of some observables.
The main objective is to investigate the underlying 
physics of the oscillatory behaviour mentioned above. 
Sometimes, physical phenomena can be more easily
 grasped with the help of simple models. Thus, the present
study may be useful to determine the mechanisms involved in this kind of  
oscillations. 

For the sake of simplicity, we consider the problem of a single 
particle moving on a one-dimensional lattice.
 At every time step, the particle can hop to
a nearest-neighbour (NN) lattice site with a probability per
unit time, which depends on the involved (initial and final) sites only.
 An additional condition is that the forward and backward hopping
rates between a given pair of NN sites must be identical. 
One of the basic ingredients of the model is a self-similar 
distribution of 
hopping rates. As we will show, this leads to a length-scale 
dependence of the diffusion coefficient that serve to explain both, 
the anomalous overall behaviour and the modulations observed in higher 
resolution measurements.

\begin{figure}[h]
\onefigure[width=\linewidth]{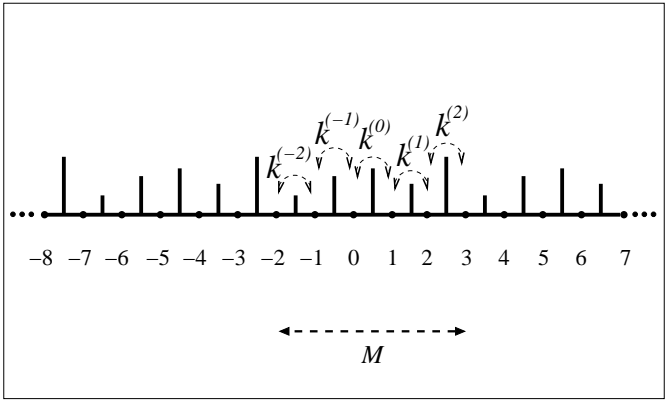}
\caption{
A one-dimensional periodic lattice with $M=5$ sites per unit
cell. A schematic barrier of height
$c/k^{(i)}$ is drawn between every pair of NN sites ($i$ and $i+1$), where
 $k^{(i)}$ is the hopping rate from site $i$ to site $i+1$ (and from site $i+1$ to site $i$). The periodicity conditions $k^{(i+M)}=k^{(i)}$ are satisfied.
}
 \label{figure1}
\end{figure}

Before going into the details of the self-similar model, let us consider
a one-dimensional periodic lattice with $M$ sites per unit cell.
The hopping rules stated above can be schematically represented by a set of 
barriers of height $h^{(i)}=c/k^{(i)}$ ($i\in Z$), where $c$ is an arbitrary 
constant and $k^{(j)}$ the probability of a hop from site $j$ to site $j+1$ 
(and from site $j+1$ to site $j$) per unit time. An sketch of this structure, 
with $M=5$, is shown in fig.~\ref{figure1}.

The effective diffusion constant $D_{\rm eff}$ for a periodic model can be
computed following the steady-state method introduced in ref.~\cite{aldao}. 
If $a$ is the distance between NN sites, it is found that

\begin{equation}
\begin{centering}
D_{\rm eff}=\left( \frac{1}{a^{2}M} 
\sum_{\begin{subarray}{l}j\; \mbox{\small in a}\\ 
\mbox{\small unit cell}
\end{subarray}}
\frac{1}{k^{(j)}}\right)^{-1}\; ,
\end{centering} 
\label{D_perio} 
\end{equation}
(in what follows we use $a=1$). The meaning of eq.~(\ref{D_perio}) is that, for 
times $t\gg t_{esc}$, where $t_{esc}$ is the average time 
for the particle to escape from the initial unit cell, the particle 
mean-square displacement $\langle  \Delta^2 x\rangle$ satisfies the normal diffusion equation

\begin{equation}
\begin{centering}
\langle\Delta^2 x(t)\rangle = \langle [x(t)-x(0)]^2\rangle=2D_{\rm eff}t\;\;\; .
\end{centering}
\label{x2_perio}
\end{equation}
This is  a very simple example of a 
length-scale dependence. There is an asymptotic regime, 
described by eq.~(\ref{x2_perio}), 
for $\displaystyle\sqrt{\langle  \Delta^2 x\rangle}$ greater than $M$, and a transitory one, 
not described here, for $\displaystyle\sqrt{\langle  \Delta^2 x\rangle}$
 smaller than $M$. 

\begin{figure}[h]
\includegraphics[width=\linewidth,clip=true]{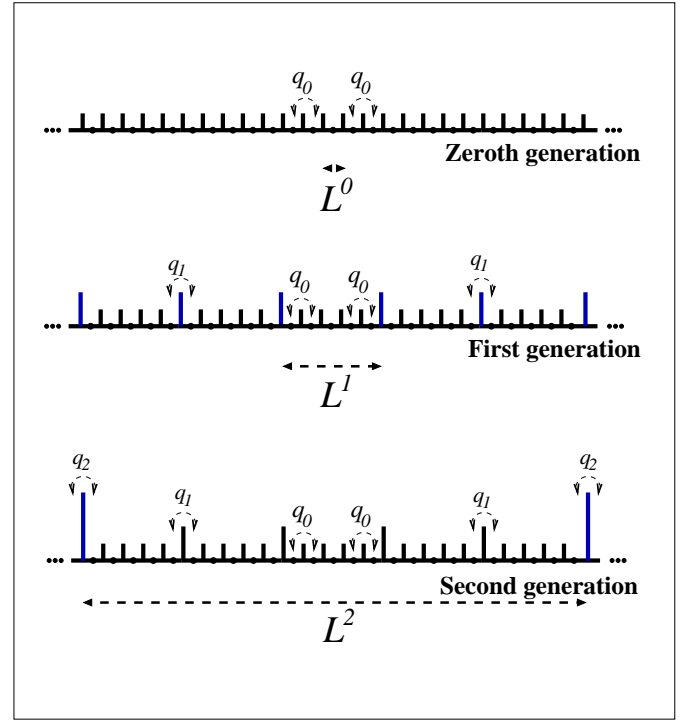}
\caption{Construction of a lattice with a self-similar distribution
of hopping rates ($L=5$). 
Top: The zeroth generation ($n=0$) with all the hopping rates
equal to $q_0$. 
Center: The first generation ($n=1$), where the 
hopping rates $q_1$ appear. The corresponding barriers are separated by a 
distance $L$. 
Bottom: The second generation ($n=2$) shows the emergence of the hopping rates
$q_2$ separated by a distance $L^2$. In the limit \mbox{$n\rightarrow\infty$}
 a self-similar distribution of hopping rates is obtained. For more details, 
see the text.
}
\label{figure2}
\end{figure}

In order to obtain the oscillations we are interested in, we proceed
now to modify the periodic lattice to allow for a greater number  of 
kinetic regimes. Let us first introduce a parameter $L$, which is an 
odd natural number greater than $1$.
The model is built in stages and the result of every stage is
called a {\it generation}. The building process is illustrated in
fig.~\ref{figure2}, for $L=5$.

The zeroth-generation lattice corresponds to the situation in which all
the hopping rates are identical 
(\mbox{$k^{(i)}=q_0$,} $\forall i\in Z$). 
In this stage, eq.~(\ref{x2_perio}),  is valid for 
$\displaystyle\sqrt{\langle  \Delta^2 x\rangle}$ greater than $1$ 
with $D_{\rm eff}=D^{(0)}$
($=q_0$, see eq.~(\ref{D_perio})).

In the first generation, the hopping rate $k^{(j)}$ is set
to $q_1$ ($<q_0$) for every  $j=pL-(L+1)/2$, with $p$ integer. 
 All the other hopping rates 
remain as in the generation zero. Eq.~(\ref{x2_perio}) is also valid in 
first-generation model but for  $\displaystyle\sqrt{\langle  \Delta^2 x\rangle}$
greater than $L$ 
and a new value of $D_{\rm eff}=D^{(1)}$ (also given by eq.~(\ref{D_perio})).

This process is iterated indefinitely and, in general, the generation $n$ is
obtained from the generation $n-1$ after replacing by $q_n$ ($<q_{n-1}$) 
the values of 
$k^{(j)}$, for every \mbox{$j=pL^n-(L^n+1)/2$}, with $p$ integer.

In the limit of an infinite number of iterations we  get the model,
with a self-similar distribution of hopping rates and an infinite
set of diffusion constants \mbox{$\{ D^{(n)}, n=0,1,2,...\}$}, discussed in the 
rest of the work.

It is not hard to convince oneself that, in this model, 
 eq.~(\ref{x2_perio}), with  $D_{\rm eff}=D^{(n)}$ , should hold for 
$\displaystyle\sqrt{\langle  \Delta^2 x\rangle}$ in some range between 
$L^n$ and $L^{n+1}$,
and this for every non-negative integer $n$.
If the RW begins from the site $j=0$ (the centre of symmetry of the lattice), 
for short enough times
(though longer than $t_{esc}^{(0)}\sim 1/q_0$) the particle behaves 
as being in the zeroth-generation lattice. It only feels the action of the 
lowest barriers and  normal diffusion, with 
a constant $D^{(0)}$, should be observed. 
However, when $\displaystyle\sqrt{\langle  \Delta^2 x\rangle}$
is of the order of $L$ ($t\sim t_{esc}^{(1)}$), the particle starts
to interact with the barriers of height $c/q_1$. For even longer length scales,
though shorter than $L^2$ ($t<t_{esc}^{(2)}$), everything happens as in the 
first-generation lattice, and one should then observe normal diffusion 
with a constant $D^{(1)}$.
Because of the self-similar properties of the lattice, this sequence of changes 
continues indefinitely, and the effective constant $D^{(n)}$,
which corresponds to the  normal diffusion in the 
\mbox{$n^{\rm th}$-generation} lattice, should appear at a scale $L^n$.

If, in addition, we impose that 
 
\begin{equation}
\begin{centering}
\frac {D^{(n)}}{D^{(n+1)}}=1+\lambda,\;\;\;\;\;\;\; \mbox{for}\;n=0,1,2,...\;\;,
\end{centering} \label{cocient}
\end{equation}
where $\lambda>0$ is another parameter of the model, both the diffusion 
coefficients and the hopping rates are determined, up to a multiplicative 
constant, (see eqs.~(\ref{D_perio}) and (\ref{cocient})) through

\begin{equation}
\begin{centering}
D^{(n)}=\frac {q_{0}}{(1+\lambda)^n},\;\;\;{\rm for}\; n=0,1,2,3,...\;\;\; ,
\end{centering}
\end{equation}
and the iterative relation 

\begin{equation}
\begin{centering}
\frac{q_{0}}{q_{i}}= \frac {q_{0}}{q_{i-1}} + (1+\lambda )^{i-1}
\lambda L^{i},\;\;\;{\rm for}\; i=1,2,3...\;\;\; .
\end{centering} \label{relation}
\end{equation}

\begin{figure}[h]
\onefigure[width=\linewidth]{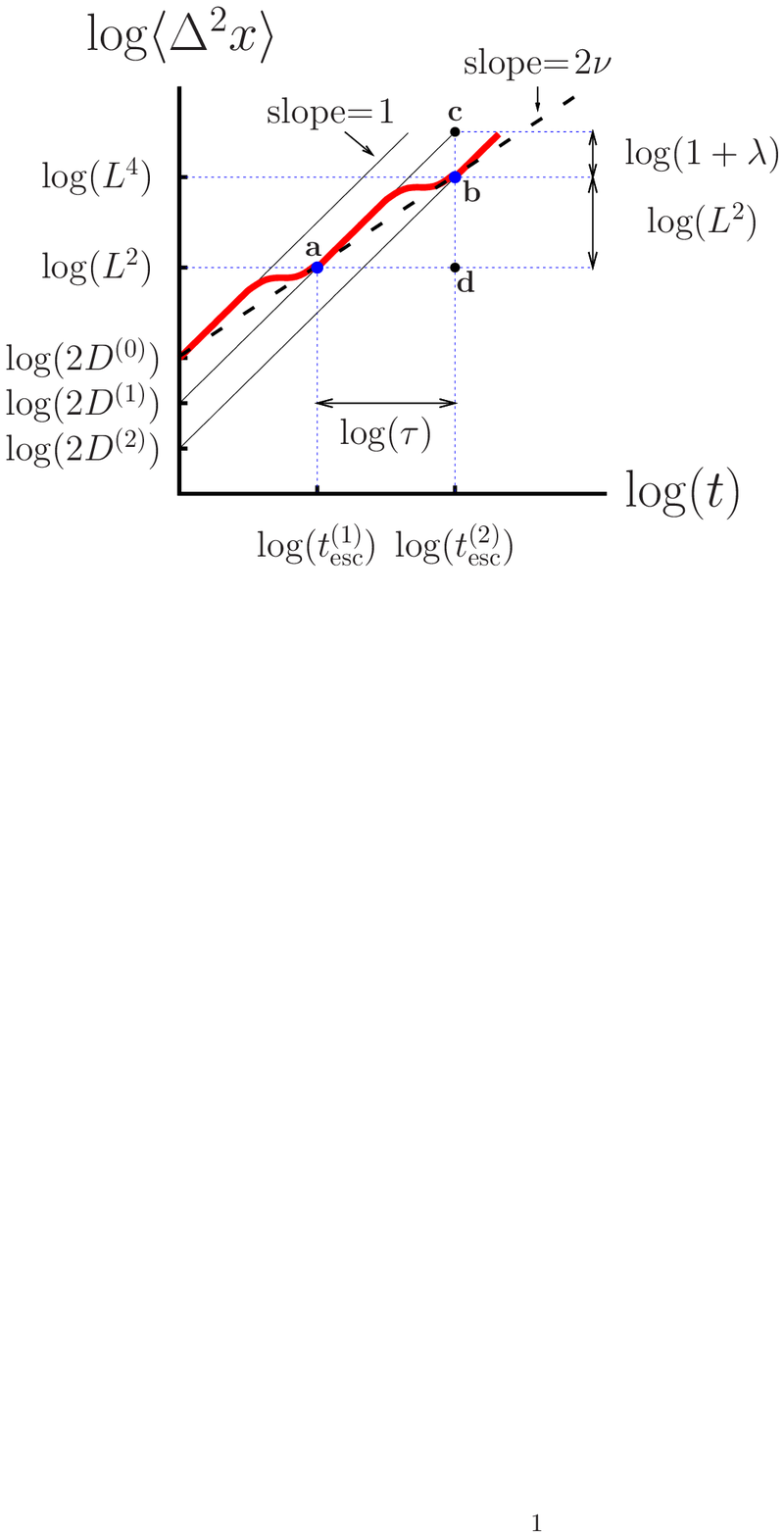}
\caption{(Color online) Schematic of the mean-square displacement
 as a function of the time, shown by the thick red curve. The length 
of the segment {\bf bc} is $\log(2D^{(1)})-\log(2D^{(2)})=\log(1+\lambda)$, 
because of eq.~(\ref{cocient}). From the slopes ($=1$) of the full straight 
lines (representing the normal diffusion behaviours, $ \langle \Delta^2x \rangle = 2 D^{(n)}t$), one gets that the segments {\bf ad} and 
{\bf cd} have the same length or, equivalently,  that 
$\log\tau=\log L^2+\log(1+\lambda)$. The dashed straight line represents the 
global power law   $ \langle \Delta^2x \rangle  \sim t^{2\nu}$, with 
$2\nu= \log L^2/\log\tau$. More details in the text. 
}
\label{figure3}
\end{figure}

From the discussion in the paragraph immediately preceding 
eq.~(\ref{cocient}), we can anticipate that
the mean-square displacement behaves qualitatively as in fig.~\ref{figure3}.
This is a sketch of $\langle\Delta^2x\rangle$ (thick curve), which, as a 
function of $t$, has a power-law form modulated by a log-periodic
amplitude. That is,

\begin{equation}
\begin{centering}
\langle \Delta^2x \rangle(t) = C t^{2 \nu} f(t),\;\;\;\;\;\;\;\;\;
 \mbox{for}\;t>t_{esq}^{(0)},
\end{centering} \label{function}
\end{equation}
where $\nu$ is the RW exponent, and $f(t)$ a 
log-periodic function, which satisfies $f(t \tau)=f(t)$, with 
the logarithmic period $\log(\tau)$.
The value of the constant $C$ is obtained by asking that the log-time average of $\log(f)$ 
over one period be zero (see, for more details, 
fig.~\ref{figure6}).
We also observe, in fig.~\ref{figure3},  two groups of inclined straight lines. 
On the one hand, the
dashed line, which corresponds to the $\langle  \Delta^2 x\rangle$ global power-law 
trend, and  has
a slope of $2\nu$. On the other hand, the solid lines, which represent
normal diffusion in each of the different generation
lattices, and have slopes of $1$. 
From these slopes, it is clear that $2\nu=\log(L^2)/\log(\tau)$ (dashed line), 
and that $\log(\tau)=\log(L^2)+\log(1+\lambda)$ (solid line), which gives
both $\tau$ and $\nu$ expressed in terms of the parameters
$L$ and $\lambda$,
 
\begin{equation}
\tau=L^{1/\nu}=(1+\lambda)L^2\;\;{\rm ,}
\label{tau}
\end{equation}

\begin{equation}
\nu = \frac{1}{2+\frac{\log(1+\lambda)}{\log L}}\;\;\;  .
\label{exponent}
\end{equation}
Note that, since $\lambda>0$, an anomalous 
diffusion appears ($\nu<1/2$, see eq.~(\ref{exponent})), and that,
from the sketch in
fig.~\ref{figure3}, we can  predict that the  amplitude of the modulation
increases with the increase of $\lambda$ or $L$.

Let us remark that the self-similarity in the mean-square displacement,
schematically shown in fig.~\ref{figure3} and mathematically described by
eq.~(\ref{function}), is a direct consequence of the set of constraints
(\ref{cocient}). Because of these relations, the distance between any pair
of nearest solid straight lines is a constant and any pair of nearest 
equivalent points in the graph (like {\bf a} and {\bf b}) are related by the transformation
($t\rightarrow\tau t$ , 
$\langle  \Delta^2 x\rangle\rightarrow\tau^{2v}\langle  \Delta^2 x\rangle$).

 Even though we have focused on the properties of the mean-square displacement,
a similar analysis applies to the average number $S(t)$ of distinct sites 
visited by the particle, after a time $t$. As we are  working
with a one-dimensional lattice, $\displaystyle S\sim\sqrt{\langle \Delta^2 x \rangle}$, and it is thus expected that 
\begin{equation}
\begin{centering}
S=C' t^{\eta}g(t)\;\;\; ,
\end{centering}
\end{equation}
where the exponent $\eta$ is equal to $\nu$, $g(t)$ is a log-periodic
function, $g(\tau t)=g(t)$ (with $\tau$ given by
eq.~(\ref{tau})) and the constant $C'$ is obtained by asking that
the log-time average of $\log(g)$ over one period be zero.

To check the validity of the analytical predictions stated above,
we have performed  Monte Carlo (MC) simulations, with $q_0=1/2$ and
a time step $\delta t=1$.
Every simulation begins with the particle at the center of a sixth-generation 
lattice  and stops after a given number of MC steps, always chosen 
to avoid that the particle reaches the highest ($6^{th}$-order) barriers.

\begin{figure}[h]
\onefigure[width=\linewidth]{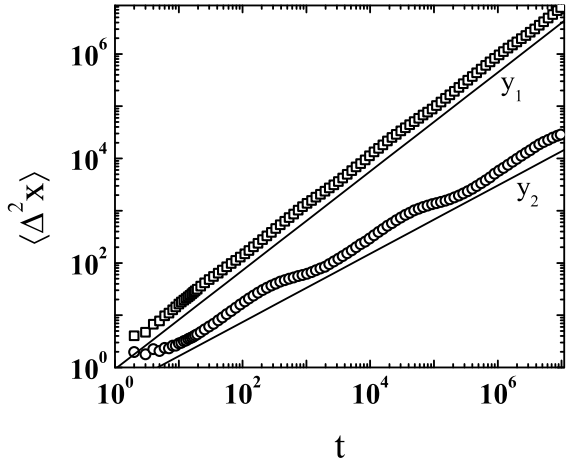}
\caption{The mean-square displacement  as a
 function of the time, for $L=5$, $\lambda =0.2$
 (squares) and $L=5$, $\lambda =5$ (circles). The straight
 lines $y_{1}$ and $y_{2}$ have slopes $2\nu$, with the values of $\nu$ 
($\nu=0.4732$ and $\nu=0.3212$, respectively),  given by
 eq.~(\ref{exponent}). These lines are drawn to guide the eyes.
}
\label{figure5}
\end{figure}

\begin{figure}[h]
\onefigure[width=\linewidth]{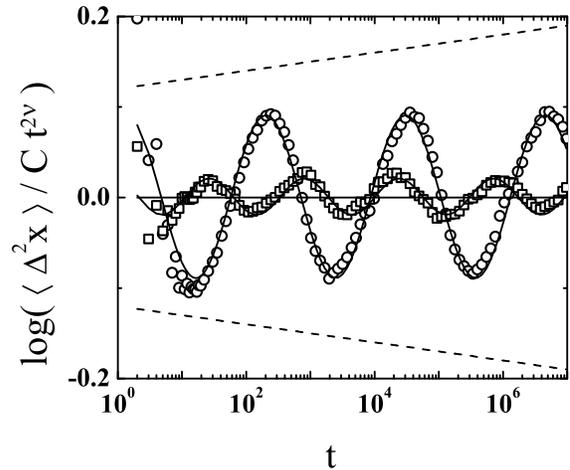}
\caption{The logarithm of the scaled mean-square displacement 
 $\displaystyle \langle
 \Delta^2 x \rangle / Ct^{2\nu}$ as a function of the logarithm of the time, 
 for $L=5$, 
 $\lambda=0.2$ (squares) and
 $L=5$, $\lambda=5$ (circles) The values of $\nu$ were obtained
 from eq.~(\ref{exponent}) and $C$ are appropriately chosen constants. 
The curvilinear lines represent first-harmonic approximations of
 the data, $\displaystyle A \sin ( (2 \pi \log t)/(\log \tau) + \alpha )$. 
The period $\tau$ is given by eq.~(\ref{tau}).  $A$ and $\alpha$ are fitted constants.}
\label{figure6}
\end{figure}

The numerical results of the mean-square displacement as a function of
the time is plotted in fig.~\ref{figure5}  for two
sets of parameters, $L=5$, $\lambda = 0.2$ and $L=5$, $\lambda=5$. It
is apparent in this figure that $\displaystyle  \langle \Delta^2 x \rangle(t)$
satisfies a modulated power law.

\begin{figure}[h]\vspace{0.2cm}
\onefigure[width=\linewidth]{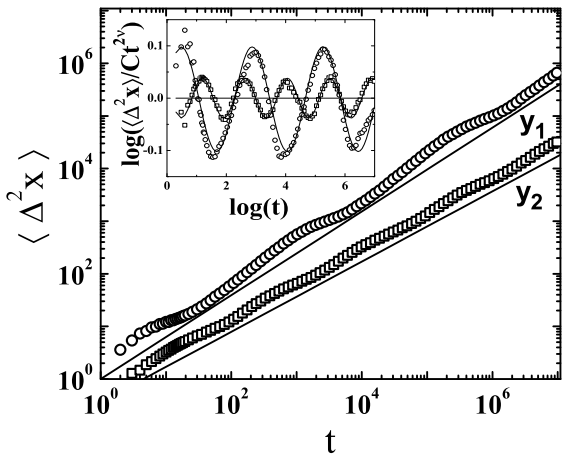}
\caption{  $\displaystyle \langle \Delta^2 x \rangle$
 against $t$ for $L=3$, $\lambda=2$ (squares) and $L=9$,
 $\lambda=2$ (circles). The straight lines $y_{1}$ and $y_{2}$
 have slopes $2\nu$, with $\nu$ ($\nu=0.40$ and
 $\nu=0.33$, respectively) obtained from eq.~(\ref{exponent}). The inset shows
 a plot of
 $\displaystyle \log \langle \Delta^2 x \rangle / Ct^{2\nu}$ vs.
 $\log t$ for the same data. The
 curvilinear lines were obtained in a similar way as in
 fig.~\ref{figure6}.}
\label{figure7}
\end{figure}

 The modulations can however be better observed in fig.~\ref{figure6},
where we have plotted  
$\displaystyle\log(\langle \Delta^2x \rangle / Ct^{2\nu})$ versus
$\log(t)$, using the same data as in  fig.~\ref{figure5}. The values of
$\nu$ were computed from eq.~(\ref{exponent}) and the constants $C$ are
appropriately chosen to have the oscillations centered around zero. 
We have drawn three straight lines. The central one 
represents the equation $\displaystyle \langle
\Delta^2 x \rangle =Ct^{2\nu}$. The other two, which have slopes of 
$2\nu\pm0.02$
and were shifted for the sake of clarity, can be used for error estimation.
The curvilinear lines are of the form 
$\displaystyle A \sin(2\pi \log(t)/\log (\tau) + \alpha )$, i.~e., the
first-harmonic approximation of a periodic function
with period $\log(\tau)$, where $A$ and
$\alpha$ are fitted parameters.  From fig.~\ref{figure6} it is indeed clear that the 
predictions in eqs.~(\ref{tau}) and (\ref{exponent}) are consistent  
with the numerical findings. It is important to emphasise that the 
theoretical-simulation agreement is 
also excellent for other set of parameters. This can, for instance, be seen in 
fig.~\ref{figure7}, where we have plotted the results of simulations for 
$L=3$, $\lambda=2$ and $L=9$, $\lambda=2$. 
Regarding the way  the amplitude of the oscillation depends on the parameters 
of the model, we can see in fig.~\ref{figure6} and in the inset of 
fig.~\ref{figure7}
how the former increases with the increase of any of the latter, which
confirm our predictions.
We would like to mention that, although not shown here, we have also
verified that the agreement between numerical and theoretical results is as 
good for the mean number of distinct visited sites as it is for the 
mean-square displacement.

\vspace{0.2cm}

\begin{figure}[h]\vspace{0.1cm}
\onefigure[width=\linewidth]{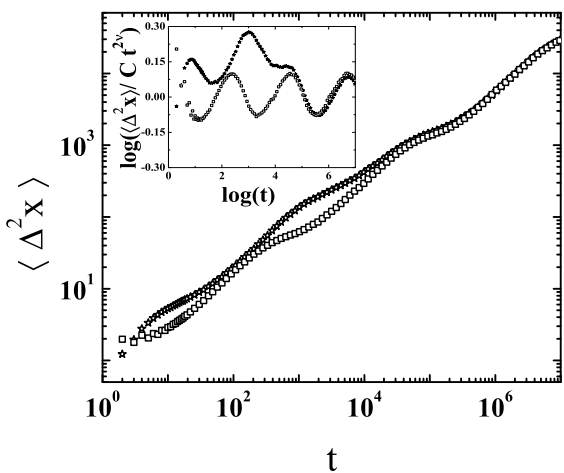}
 \caption{Dependence of the mean-square displacement  on the
 initial position. The squares represent the time behaviour
 of RW's that begin at site $i=0$. The stars, those that begin
 at site $i=12$.
  The inset shows $\log(\langle \Delta^2 x\rangle /Ct^{2 \nu})$ versus $\log t$ for the same data.
 }
\label{figure9}
 \end{figure}

So far, we have considered RW's that always begin from the centre of symmetry
(central site in fig.~\ref{figure2}). To examine 
what happens when the RW begins from another site, 
the evolution of the mean-square displacement 
on a lattice with $L=5$  and $\lambda=5$ is plotted, in fig.~\ref{figure9}, for two different 
initial conditions. In one sample (squares), the particle was initially
located at $j=0$ (the center of symmetry), in the other (stars), at  
$j=12$ (the rightmost site in \mbox{fig.~\ref{figure2}-bottom}). 
As can be expected, two clearly different behaviours are present at
short times. However, at long times, a superposition of the data is 
observed. This can be understood as
follows. For times much longer than the time for the
particle to escape the second-generation unit cell
(i.e., when
$\displaystyle\sqrt{\langle \Delta^2 x\rangle}\gg L^{2}$) the diffusion
is governed by the coefficients $D^{(n)}$, with $n \geq
2$, and becomes independent of the initial position. Similar
results were obtained for several initial positions and
different values of {\it L} and $\lambda$. In general, it can be said that 
for $\displaystyle\sqrt{\langle \Delta^2 x\rangle} \gg L^{n}$, 
the behaviour of the RW does not dependent of the initial position, provided that 
the latter is at a distance less than $(L^{n}-1)/2$ from the centre of symmetry.

To summarise, we have presented a one-dimensional 
self-similar model of RW, which yields a log-periodic modulated
power law for the most commonly measured quantities 
($\langle  \Delta^2 x\rangle(t)$ and $S(t)$). 
The goal of this paper it to introduce a minimal model
that provides a simple way to gain insight into the effect of
self-similarity on this kind of oscillations. As it was mentioned above, 
log-periodic modulation is characteristic 
of fractals with finite ramification order \cite{gra,kro,ace,bab1,bab2}.
In these objects, at a given length-scale, a
random walker cannot easily pass from one array to the next, due to the
small number of connections between equivalent structures;
 a difficulty that increases with the structure linear size (because of the
larger ratio of internal to connecting sites).
 In the model here, a similar effect occurs at a length
$L^n$, as a consequence of the
 decrease of the hopping rates at the edges
 of the $n^{th}-$generation unit cell, when $n$ changes
 to $n+1$. This may be considered as a rough qualitative explanation of the
 log-periodic modulations in fractals with finite connexion order. 
 Even when, for these objects, 
 we have not calculated how the effective diffusion
 constant depends on a characteristic length, we expect that
 the knowledge of this dependence would allow us to obtain the values
 of $\nu$ and $\tau$, in the same way as for the one-dimensional model. 
 Finally, let us stress that, by tuning the values of the two parameters
 $L$ and $\lambda$, it is possible to  
 design a structure that leads to an oscillatory power-law behaviour
 with predetermined values of $\nu$ and $\tau$ (whenever $\nu<1/2$ and
$\tau= L^{1/\nu}$). So far, the parameter $L$ has been restricted to odd
values, because in these cases the lattice has a centre of symmetry. 
However, as 
fig.~\ref{figure9} shows, the initial position of the random walker is not 
relevant for its long-time behaviour and thus the model can be easily 
generalised to include even  values of $L$ greater than $2$.

\acknowledgments
One of us (H.O.M) is grateful to E.~V.~Albano for interesting
discussions and for kindly providing a copy of ref.~\cite{bab2} before 
publication. We thank F. Rom\'a and S. Bustingorry for useful literature.
This work was supported by the Universidad Nacional de Mar del Plata, the 
Consejo Nacional de Investigaciones Cient\'{\i}ficas y T\'ecnicas --CONICET-- 
(PIP-5648) and the Agencia Nacional de Promoci\'on Cient\'{\i}fica y 
Tecnol\'ogica --ANPCyT-- (PICT 2004 17-20075).

\end{document}